\documentstyle[12pt,epsf]{article}

\begin{document}
\baselineskip 24pt

\title{Coupled Map Modeling for Cloud Dynamics\\}

\author{Tatsuo Yanagita
     \thanks{ email address:{\tt yanagita@elsip.hokudai.ac.jp} }\\
     \small \sl Institute of Electronic Science,\\
     \small \sl Hokkaido University, Sapporo, Hokkaido 060, Japan\\
     \\     and\\
     \and
Kunihiko Kaneko
     \thanks{ email address:{\tt kaneko@complex.c.u-tokyo.ac.jp} }\\
     \small \sl Department of Pure and Applied Sciences,\\
     \small \sl University of Tokyo, Komaba, Meguro-ku, Tokyo 153, Japan\\
}

\date{}

\maketitle

\begin{abstract}
\baselineskip 24pt
A coupled map model for cloud dynamics is proposed, which 
consists of the successive operations of the
physical processes; buoyancy, diffusion, 
viscosity, adiabatic expansion, fall of a droplet by gravity, 
descent flow dragged by the falling droplet, and advection. 
Through extensive simulations, the phases corresponding to
stratus, cumulus, stratocumulus and cumulonimbus are found, with the
change of the ground
temperature and the moisture of the air.  They are
characterized by order parameters such as the cluster number,
perimeter-to-area ratio of a cloud, and Kolmogorov-Sinai entropy.
\end{abstract}

PACS\#
47.52.+j,
92.40.Cy,
92.60.Jq

Cloud dynamics plays
an important role in the climate system, weather forecast, geophysics 
and so on. 
However this elementary process in meteorology is very much complicated 
because it consists of different time and space 
scales and the phase transition from liquid to gas is 
coupled with the motion of atmosphere.
Even if the flow of the atmosphere were known with accuracy using 
Navier-Stokes equation, we could not discuss the morphology of cloud.
In order to investigate such a complex system, construction of a
phenomenological model is essential.
In this letter, we introduce a coupled map lattice model of cloud formation 
which reproduces the diversity of cloud patterns.   Characterizations of four 
cloud phases are also given.

Coupled map lattices (CML) are useful to study the dynamics of spatially 
extended systems \cite{KANEKO5}. 
Recently, CML has successfully been applied to spinodal decomposition \cite{OONO2}, 
Rayleigh-B\'{e}nard convection \cite{YANAGITA4-2}, the boiling 
transition \cite{YANAGITA3-2}, 
and so on\cite{CHAOS1}. 

Here we construct a CML model of cloud in a 2-dimensional space. 
CML modeling is based on the separation and successive operation of procedures,
which are represented as maps acting on a field variable on a lattice \cite{KANEKO2}. 
Here, we choose a two dimensional square lattice $(x,y)$ with $y$ as 
a perpendicular direction, and assign the velocity field $\vec{v}^t (x,y)$, 
the mass of the vapor $w_v^t(x,y)$ and the liquid $w_{\ell}^t(x,y)$, and the 
internal energy $E^t (x,y)$ as field variables at time $t$. 
The dynamics of these field variables consists of Lagrangian and Eulerian parts. 
For the latter part, we adopt the following processes 
\footnote{Here we consider such spatial scale that the 
centrifugal and Coriolis forces are neglected.};
(1) heat diffusion (2) viscosity
(3) buoyance force (4) the pressure term requiring $div \vec{v}$ to be 0, in 
an incompressible fluid \footnote{If the vertical air motion is 
confined within a shallow layer, the motion of atmosphere can be regarded as 
incompressible flow \cite{LANDAU2-1}.}.
Here, we use the discrete version of $grad(div \vec{v})$ which refrains from 
the growth of $div \vec{v}$. Indeed, we have already constructed
the CML representations of  the above four procedures 
\cite{YANAGITA4-2}that agree
with experiments on the Rayleigh-B\'{e}nard convection,
which is a cardinal role for cloud dynamics.
(5) diffusion of vapor $w_v^t(x,y)$;
(6) adiabatic expansion; assuming the adiabatic process and the equilibrium 
ideal gas with gravity field, we adopt such an approximation that
the temperature of the parcel risen from the
height $y$ to $y+\Delta y$ is 
decreased in proportion to the displacement $\Delta y$. 
Thus the temperature $E^t(x,y)$ is decreased in proportion to $v_y^t(x,y)$. 
(7) phase transition from vapor $w_v^t(x,y)$ to liquid $w_{\ell}^t(x,y)$  and vise 
versa accompanied by the latent heat.
Here, we use  the simplest type of the bulk water-continuity model in 
meteorology \cite{HOUZE}.
The dynamics is represented as a relaxation to an equilibrium 
point $w^*(x,y)$ which is a function of temperature $E^t(x,y)$.
(8) the dragging force; assuming that the droplets are uniform in size and 
fall with the terminal velocity $V$ with  neglect of the relaxation time to it,
the dragging force is proportional to the product of the 
relative velocity $(v_y^t(x,y)-V)$ and the density of droplets $w_{\ell}^t(x,y)$ 
at a lattice site.

Combining these dynamics, the Eulerian part is written as the successive operations of the following mappings (hereafter we use the notation for 
discrete Laplacian operator: $\Delta A(x,y)=\frac{1}{4}\{ A(x-1,y)+A(x+1,y)+A(x,y-1)+A(x,y+1)-4A(x,y)\} $ for any field variable $A(x,y)$):
For convenience, we represent state variables after an operation of each 
procedure with the superscript $t+1/n$ where $n$ is the total number of
procedures.

{\bf Buoyancy and dragging force}
\[
v_y^{t+1/3}(x,y)=v_y^t(x,y)+\frac{c}{2}\{E^t(x+1,y)+E^t(x-1,y)-2E^t(x,y) \}
-\gamma w_{\ell}(x,y)(v_y(x,y)-V)
\]

{\bf Viscosity and pressure effect}
\begin{eqnarray}
\vec{v}^{t+2/3}(x,y) \nonumber
& = & \vec{v}^{t+1/3}(x,y)+\nu \Delta \vec{v}^{t+1/3}(x,y) + \eta grad (div \vec{v}^{t+1/3}(x,y))
\end{eqnarray}
with $grad (div \vec{v})$ as its discrete representation on the lattice \cite{YANAGITA4-2}.

{\bf Thermal diffusion and adiabatic expansion}
\[
E^{t+1/3}(x,y)= E^{t}(x,y) + \lambda \Delta E^t(x,y)-\beta v_y^{t}(x,y)
\]

{\bf Diffusion of vapor}
\[
w_v^{t+1/3}(x,y))=w_v^{t}(x,y)+\lambda \Delta w_v^t(x,y)
\]

{\bf Phase Transition}
To get the procedure, we use the
discretized version of the following linear equations 
for the relaxation to equilibrium point $w^*$:
\begin{eqnarray}
\frac{d w_v(x,y)}{dt} & = & +\alpha (w_v(x,y)-w^{*}) \nonumber \\
\frac{d w_{\ell}(x,y)}{dt} & = & -\alpha (w_v(x,y)-w^{*})  \nonumber \\
\frac{d E(x,y)  }{dt} & = & -Q(\frac{d w_v(x,y)}{dt}-\frac{d w_{\ell}(x,y)}{dt}) \nonumber  
\end{eqnarray}

\[
w^{*}=\left\{
\begin{array}{ll}
A \exp(q/(E+const.)) & \mbox{ if $>W(x,y)$} \\
W(x,y)      & \mbox{otherwise}
\end{array}
\right.
\]

which form is chosen to be consistent with the Clausius-Clapeyron's
equation exp(-q/Temperature) \cite{LANDAU3-1},
while $W(x,y)=w_{\ell}(x,y)+w_v(x,y)$ is the total mass of water.

The Lagrangian scheme expresses the advection of velocity, temperature, liquid and 
vapor.  This process is expressed by the motion of
a quasi-particle on each lattice site 
$(x,y)$ with velocity $\vec{v}(x,y)$.  We adopt the method
presented in \cite{YANAGITA4-2}, while for the liquid variable $w_{\ell}(x,y)$,
we also include the fall of a droplet
with a final speed $V$. Thus, the quasi-particle moves
to $(x+v_x(x,y), y+v_y(x,y)-V)$ to allocate $w_{\ell}(x,y)$ at its 
neighbors. 
Through this Lagrangian procedure, the energy and momentum are conserved.

Summing up, our dynamics is given by successive 
applications of the following step;

\[
\left \{ \begin{array}{c}
           \vec{v}	^t		(x,y) \\
           E		^t		(x,y) \\
           w		^t		(x,y) 
          \end{array}
\right \}
\stackrel{\rm  {Buoyancy+Dragging}}{\longmapsto}
\left \{ \begin{array}{c}
           \vec{v}	^{t+1/3}	(x,y) \\
           E		^t		(x,y) \\
           w		^t		(x,y) 
          \end{array}
\right \}
\stackrel{\rm {Viscosity+Pressure}}{\longmapsto}
\left \{ \begin{array}{c}
           \vec{v}	^{t+2/3}	(x,y) \\
           E		^t		(x,y) \\
           w		^t		(x,y) 
          \end{array}
\right \}
\stackrel{\rm {Diffusion}}{\longmapsto}
\]
\[
\left \{ \begin{array}{c}
           \vec{v}	^{t+2/3}	(x,y) \\
           E		^{t+1/3}	(x,y) \\
           w		^{t+1/3}	(x,y) 
          \end{array}
\right \}
\stackrel{\rm {Phase Transition}}{\longmapsto}
\left \{ \begin{array}{c}
           \vec{v}	^{t+2/3}	(x,y) \\
           E		^{t+2/3}	(x,y) \\
           w		^{t+2/3}	(x,y) 
          \end{array}
\right \}
\stackrel{\rm {Advection+Gravity}}{\longmapsto}
\left \{ \begin{array}{c}
           \vec{v}	^{t+1}		(x,y) \\
           E		^{t+1}		(x,y) \\
           w		^{t+1}		(x,y) 
          \end{array}
\right \}
\]

For the boundary, we choose the following conditions; 
(1) Bottom plates: Assuming the correspondence between $E$ and the temperature, we choose $E(x,0)=E_0$. 
(2) Top plates: We choose the no-flux condition 
$E_{t}(x,N_y)-E_{t}(x,N_y-1)=0$.
For both the plates, we choose the no-slip condition for the velocity field,
and adopt the reflection boundary for the Lagrangian procedure.
The liquid and vapor  are fixed at zero for both the plates.
(3) Side walls at $x=0$ and $x=N_x$: We use periodic boundary conditions.

The basic parameters in our model are the temperature $E_0$ at the ground, 
the Prandtl number (ratio of viscosity to heat diffusion $\nu/\lambda$), 
adiabatic expansion rate $\beta$, the terminal velocity of liquid droplets $V$, 
the coefficient for the dragging force $\gamma$, the phase transition rate 
$\alpha$, the latent heat $Q$  and the aspect ratio ($N_x/N_y$). 
Hereafter we fix these parameters as 
$\lambda=0.2,\eta=\nu=0.2,\beta=0.2,V=0.2,\gamma=0.2,\alpha=0.2,Q=0.2$, and
study the change of the morphology in cloud as the ground temperature $E_0$ and the 
total mass of water $W=\sum_{x,y}(w_v(x,y)+w_{\ell}(x,y))$ are varied
(note that  $W$ is conserved).

\begin{figure}[htbp]
\begin{center}
{\def\epsfsize#1#2{0.6#1}\epsfbox{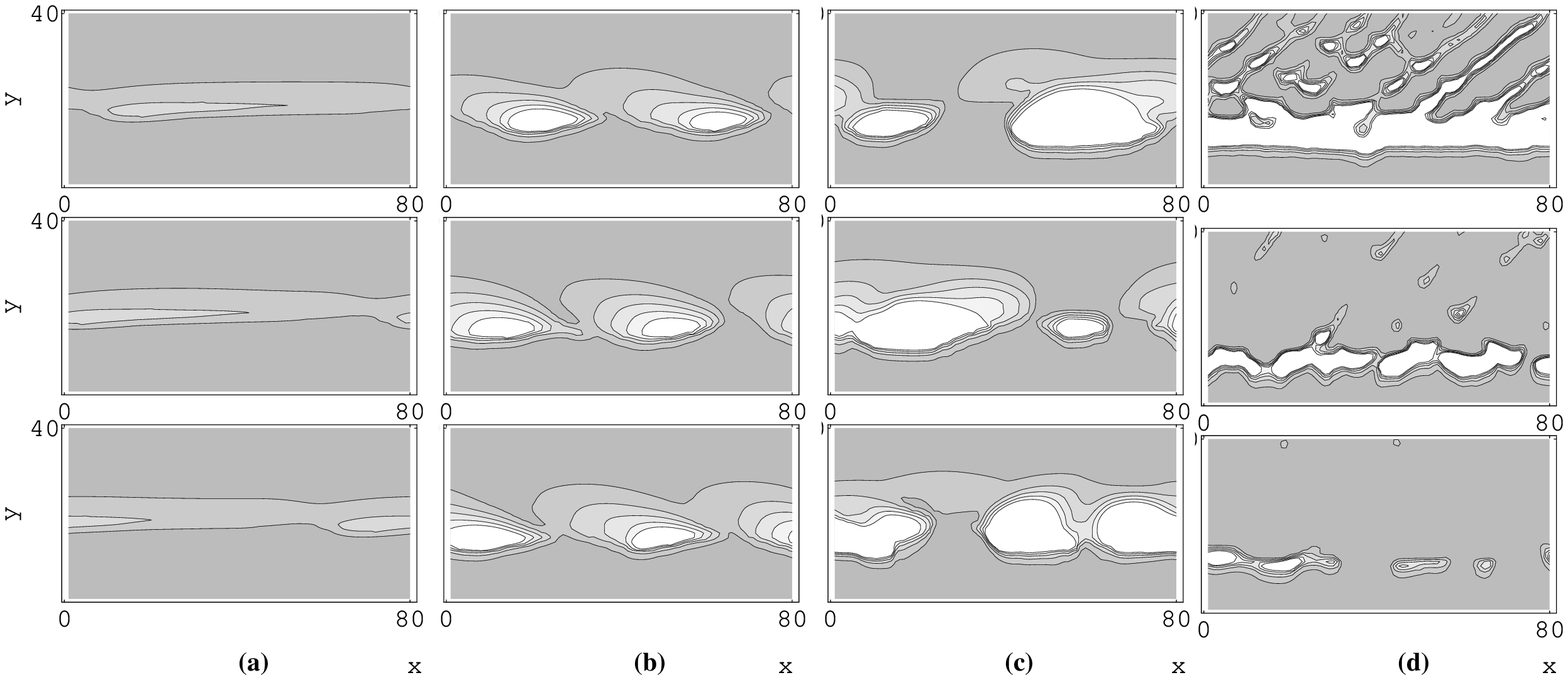}}
\end{center}
\vspace{-0mm}
\caption{}
\begin{quote}
\normalsize
Temporal evolution of cloud patterns. 
Snapshot of the mass of the liquid $w_l^t(x,y)$ is shown with the use of gray 
scale. The darker the pixel, the lower the liquid is.
In other words, the white region corresponds to the
high density of liquids, i.e., to a cloud region.
Snapshot patterns are plotted from top to down per 200 steps after 
initial 5000 steps of transients, starting from a random initial condition.
(a) Stratus ($E_0=3.0,W=0.006$),
(b) Cumulus ($E_0=3.0,W=0.007$),
(c) Cumulonimbus ($E_0=4.0,W=0.009$),
(d) Stratocumulus($E_0=5.0,W=0.009$).
The lattice size is $(N_x, N_y)=80 \times 40$.
\end{quote}
\label{fig:pattern2d}
\end{figure}

To see the spatiotemporal dynamics, the evolution of the mass of the liquid 
$w_{\ell}(x,y)$ is studied.
In Fig.~\ref{fig:pattern2d}, four typical time evolutions of $w_{\ell}(x,y)$ are plotted.
By changing $E_0$ and 
$W$, the following four types of cloud have been  found; 
(a) ``stratus'', (b) ``cumulus'', (c) `` stratocumulus'' and (d) ``cumulonimbus''.

\begin{figure}[htbp]
\begin{center}
{\def\epsfsize#1#2{0.8#1}\epsfbox{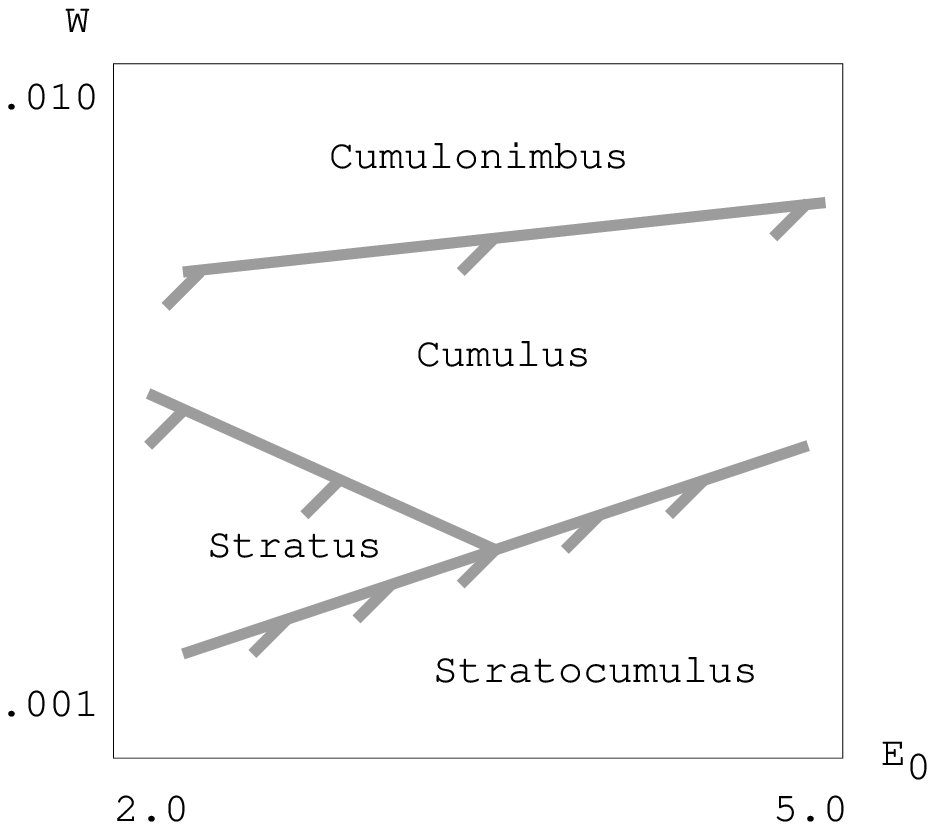}}
\end{center}
\caption{}
\vspace{-0mm}
\begin{quote}
\normalsize
Phase diagram for the morphology of cloud.
The term stratus, cumulus, cumulonimbus and stratocumulus correspond to the
pattern (a),(b),(c) and (d) in Fig.~\ref{fig:pattern2d}, respectively.
\end{quote}
\label{fig:schematic}
\end{figure}

Stratus is a thin layered pattern of cloud,  while cumulus is a
thick lump of cloud.  These two patterns are rather stable, while
the other two patterns are dynamically unstable.
At stratocumulus, 
a thin layered cloud pattern is torn into pieces and 
small fragments of clouds are scattered.
These scattered clouds vanish while a new layered cloud is formed again later.
The formation and annihilation of clouds are periodically
repeated.
Cumulonimbus is a thicker cloud in height than cumulus.  Besides
the size change, the cloud pattern is unstable. The clouds split and 
coalesce repeatedly.
The classification into four types is based on the comparison between
our spatiotemporal pattern and the definition by meteorology \cite{HOUZE}, while
the phase diagram is given in Fig.~\ref{fig:schematic},
which is obtained from the pattern and quantifiers to be discussed.  
Summarizing the diagram,
a cumulus or cumulonimbus is observed under the condition of rich moist air 
while a stratus appears in small $W$ and under low temperature.

\begin{figure}[htbp]
\begin{center}
{\def\epsfsize#1#2{0.7#1}\epsfbox{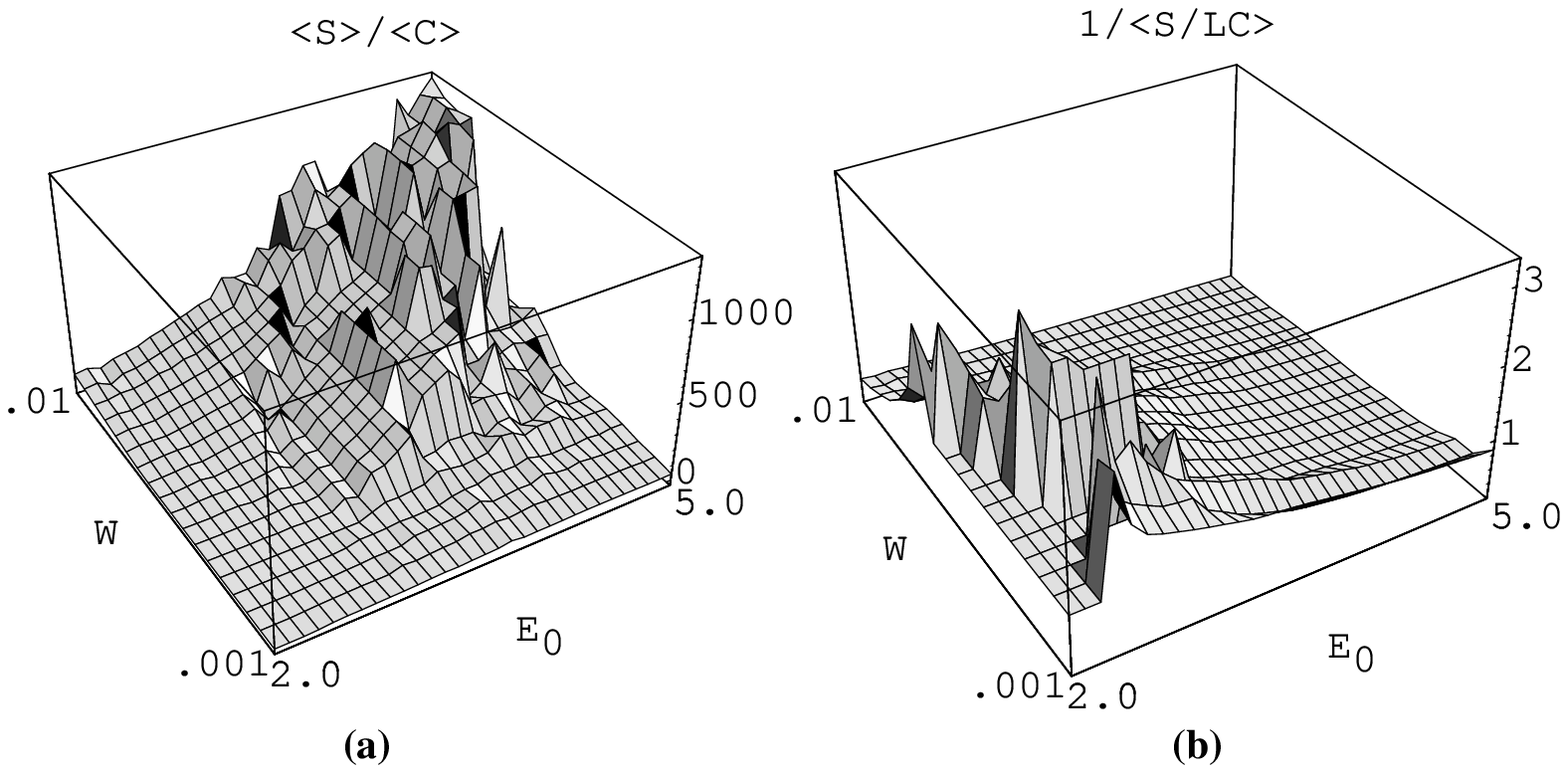}}
\end{center}
\caption{}
\begin{quote}
\normalsize
\normalsize
(a) The average size of a cloud cluster $\langle S \rangle/\langle C \rangle$
and (b) the stratus order parameter (SOP),
plotted with the change of  $W$ and $E_0$.  The
average is taken over 20000 time steps after discarding 
10000 steps of transients, starting from a random initial condition.
\end{quote}
\label{fig:diagram}
\end{figure}

To classify these patterns quantitatively, we have measured
several order parameters.  First, we
define a cloud cluster as connected lattice sites in which 
$w_{\ell}(x,y)$ is larger than a given threshold $w_c$. 
\footnote{The phase diagram of the morphology of cloud does not depend on 
the choice of the threshold $w_c$ if $0.02<w_c<0.04$.}
$C(t)$ is defined as the number
of clusters disconnected with each other.
Then, the ``cloudiness''
is measured by the total number of cloud sites, that is,
$S(t)=\sum_{x=1}^{N_x} \sum_{y=1}^{N_y} \Theta(w_{\ell}(x,y)-w_c) $,
where $\Theta(x)$ is Heavisede function.
The (temporal) average of the cluster number is large at the onset of 
cloud formation (i.e., small
$E_0$ and $W$), and at the stratocumulus.
The quantity $\langle S \rangle/\langle C \rangle$ measures
the average size of each cloud cluster.  It
is larger at cumulus and is largest at cumulonimbus 
(Fig.~\ref{fig:diagram}-(a)).

To characterize the difference between stratus and cumulus,
the morphology of clouds should be taken into account.  
Roughly speaking, the
stratus is a one-dimensional like pattern while the
cumulus is a two-dimensional one.  To see the morphological difference,
we have measured the total perimeter of 
cloud $L(t)=\sum_{x=1,y=1}^{N_x,N_y} \sum_{\delta x=\pm 1, \delta y=\pm 1} 
\Theta(w_{\ell}(x,y)-w_c) \Theta(w_c-w_{\ell}(x+\delta x,y+\delta y))$.
``Stratus order parameter'' (SOP) is introduced as 
$\langle 1/(S(t)/(L(t)C(t)))\rangle_t$, the inverse
of the ratio of area to perimeter per cluster.
If it is large the pattern is close to a one-dimensional object.
Change of SOP with $E_0$ and $W$ is plotted in Fig.~\ref{fig:diagram}(b).
As is expected by the definition of SOP and
the thin nature of stratus cloud,
it has a larger value at the stratus phase and takes a lower value
at the cumulus.

Of course, dynamical quantifiers are important to characterize the cloud 
patterns.  For example, the fluctuations of the cluster size, 
SOP, and $S/C$ are larger at the cumulonimbus and stratocumulus phases.
To see the dynamics closely, we
have also measured the time series of the
spatial sum of the mass of the liquid $\overline{L(t)}=\sum_{x,y} w_{\ell}^t(x,y)$, 
which corresponds to the cloudiness of the total space.
The evolution of $\overline{L(t)}$ is 
almost stationary at stratus and cumulus, with only tiny fluctuations.
The change is periodic at stratocumulus and chaotic at cumulus.
It should be noted that the low-dimensional
collective dynamics of the total liquid emerges even if
the spatiotemporal dynamics is high-dimensional chaos.

\begin{figure}[htbp]
\begin{center}
{\def\epsfsize#1#2{0.7#1}\epsfbox{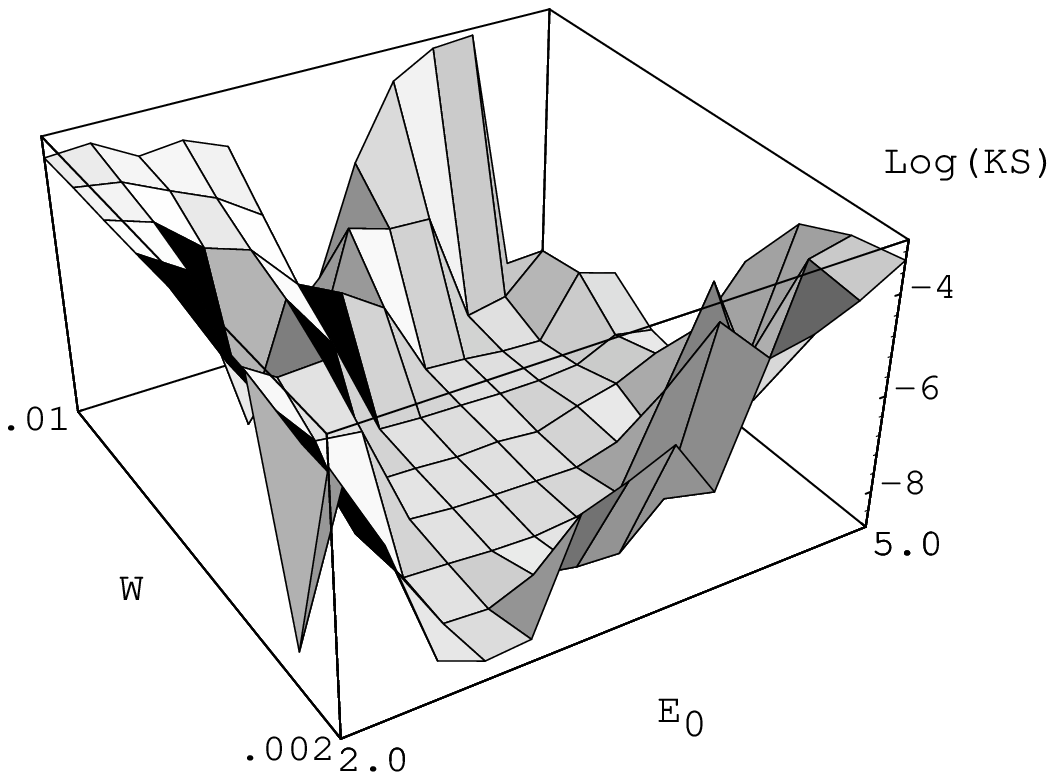}}
\end{center}
\caption{}
\begin{quote}
\normalsize
KS entropy calculated by the sum of positive Lyapunov exponents is
plotted versus $W$ and $E_0$. 
The first 20 Lyapunov exponents are computed by averaging over 
20000 time steps after discarding initial 5000 steps. 
\end{quote}
\label{fig:KS}
\end{figure}

To characterize chaotic dynamics, Kolmogorov-Sinai(KS) entropy is estimated by
the sum of positive Lyapunov exponents, as is plotted in Fig.~\ref{fig:KS}(b). 
It has a larger value at stratocumulus and cumulonimbus, which
implies that the cloud dynamics there is chaotic both spatially and temporally.
It is also positive at a lower temperature 
that corresponds to the onset of cloud formation,
where the dynamics is unstable.

In summary, we have proposed a CML model for pattern formation of cloud by introducing a simple phase transition dynamics from liquid to vapor, so called 
the bulk water-continuity model \cite{HOUZE}.
Our model reproduces the diversity of cloud patterns: 
stratus, stratocumulus, cumulus and cumulonimbus. 
This agreement implies that the qualitative feature of cloud dynamics
is independent of microscopic details such as
detailed droplet formation dynamics.

In order to globally understand the phenomenology, 
the present computationally efficient model is powerful, which makes us
possible to characterize the cloud phases.
We believe that the observed phase diagram for the morphology of the cloud
is valid for the cloud in nature. It is also interesting to
propose that similar phase changes may be seen generally in
convective dynamics including phase transitions,
since detailed processes specific only to clouds are abstracted
in our model.
Extensions of the present model to a three-dimensional case and
inclusions of centrifugal and Coriolis forces are rather straightforward.
By these extensions, study of the global atmosphere dynamics 
of a planet will be possible.

This work is partially supported by Grant-in-Aids for Scientific Research from the Ministry of Education, Science, and Culture of Japan, and by a cooperative research program at Institute for Statistical Mathematics.

\end{document}